# Exceptional Anti-Icing Performance of Self-Impregnating Slippery Surfaces

*Christos Stamatopoulos,*[†] *Jaroslav Hemrle,*[‡] *Danhong Wang,*[†] *Dimos Poulikakos*[†,*]

[†] Laboratory of Thermodynamics in Emerging Technologies, Mechanical and Process Engineering Department, ETH Zürich, Sonneggstrasse 3, 8092 Zurich, Switzerland

[‡] ABB Switzerland, Corporate Research, Segelhofstrasse 1K , 5405 Baden-Daetwill, Switzerland

**ABSTRACT:** A heat exchange interface at subzero temperature in a water vapor environment, exhibits high probability of frost formation due to freezing condensation, a factor that markedly decreases the heat transfer efficacy due to the considerable thermal resistance of ice. Here we report a novel strategy to delay ice nucleation on these types of solid-water vapor interfaces. With a process-driven mechanism, a self-generated liquid intervening layer immiscible to water, is deposited on a textured superhydrophobic surface and acts as a barrier between the water vapor and the solid substrate. This liquid layer imparts remarkable slippery conditions resulting in high mobility of condensing water droplets. A large increase of the ensuing ice coverage time is shown compared to the cases of standard smooth hydrophilic or textured superhydrophobic surfaces. During deicing of these self-impregnating surfaces we show an impressive tendency of ice fragments to skate expediting defrosting. Robustness of such surfaces is also demonstrated by operating them under subcooling for at least 490hr without a marked degradation. This is attributed to the presence of the liquid intervening layer, which protects the substrate from hydrolyzation enhancing longevity and sustaining heat transfer efficiency.

## 1. INTRODUCTION

Inhibition and control of ice nucleation and low ice adhesion play a critical role to a broad range of daily life and industrial applications[1] such as power transmission (high-voltage lines and towers),[2] aerodynamics (aircrafts, helicopters and wind turbines)[3] and heat exchangers (refrigerators).[4,5] Even in applications where icing is a process that has to be promoted i.e. ice generators[6,7] and electrothermal energy storage systems,[8,9] the surfaces and interfaces through which heat exchange occurs in order to freeze water, is necessary to be maintained ice-free through continuous ice removal. Otherwise, ice production becomes inefficient due to the added high thermal resistance of the formed ice layer.

Icing challenges that are being explored by the research community and are closely related to the aforementioned applications, are the processes of condensate freezing[10] and frosting due to desublimation[5] escorted by ice propagation[11] and considerable ice adhesion.[12,13] Final target is the design of icephobic surfaces that can be employed in real applications.

A conventional strategy for constricting icing is the exploitation of hydrophobicity by applying low surface energy coatings[12] on smooth surfaces. Based on these coatings (usually made of a polymer), it was shown that, apart from increasing the critical activation energy for heterogeneous nucleation with the increase in the static contact angle,[14,15] diminishing of ice adhesion can be also correlated with an increase in the receding contact angle.[12] Such coatings must be applied periodically, since they shed with time. They are also limited in their ice adhesion strength performance, since their receding contact angle intrinsically cannot exceed approximately $\theta_{rec}=120^{o}$.[16,17] However, very recently it was shown that polymeric coatings can be rationally engineered and become a novel and reliable solution in the promotion of robust icephobicity. In particular, by modifying the cross link density of a polymer its interfacial slippage can be regulated so that it exhibits remarkable low ice adhesion strength.[18]

A different more advanced strategy, which addresses the aforementioned issues of an intrinsically icephobic surface, is strongly related to the employment of an intervening layer between the water vapor and the solid phase, which can act as a barrier. Specifically, by forming a vapor[19–24] or liquid[25–29] anti-adsorptive layer between the solid substrate and the water vapor, the diffusion process of water molecules in the respective layer is hindered, preventing penetration of water into the asperities of the solid surface. Consequently, the initiation of ice nucleation will be limited to the liquid-vapor interface (desublimation) or the liquid-liquid interface (condensate freezing).

For the case of a gaseous intervening layer, it was shown that by forming a vapor layer of a non-condensable gas (carbon dioxide) originating from a sublimating carbon dioxide substrate under ultra-low temperature (i.e. -79 $^{o}$C ), the impacting water droplet can rebound without solid contact, preventing it from freezing.[22] Similarly, a lotus type superhydrophobic surface with a micro to nano scale hierarchical roughness, forms air pockets inside its surface textures minimizing the capillary forces between the solid substrate and the droplet. Consequently, a droplet can easily slip or rebound before freezing occurs.[30] However, microtextured superhydrophobic surfaces can lose their high water repellency property and icephobic characteristics, since capillary condensation with subsequent freezing may occur indiscriminately without any particular spatial preference, resulting in ice adhesion increase compared to standard smooth surfaces and ice accumulation.[31,32] Even if this formed ice layer were to be melted, the resulting liquid would be in fully impaled Wenzel state rendering the process of dewetting with transition to Cassie-Baxter state a very difficult task even when external (such as mechanical) force is exerted.[33] In contrast to microstructuring, it has been recently indicated that nanotexturing[34,35] can potentially become a highly competitive anti icing strategy. It was shown that it can effectively prevent icing from penetrating the nanoscale features of the surface as a consequence of the stable entrapment of air pockets inside its nanoroughness. Due to this effect, the defrosting process becomes facile, since frost transforms into liquid drops in Cassie-Baxter state that easily depart from the nanotextured superhydrophobic surface under the effect of gravity and low inclination angles ($<15°$).[34]

An alternative, potentially effective anti-icing approach consists of employing anti-adsorption process, with the formation of a liquid layer with low surface tension and negligible solubility in water, impregnating the surface roughness. With this approach a stable, ultra-smooth, liquid layer is infused into a micro/nano structured and chemically functionalized surface texture. An additional advantage of this strategy is the inherently reduced impalement risk of liquid water or ice into the texture, due to the fact that the infused liquid layer is much more difficult to be displaced than a gas layer. Although very promising results were shown in terms of anti-icing performance[29] of these kind of surfaces (LIS, SLIPS),[27,29,36–38] it has been found that the process of subcooled condensate freezing on such lubricant-impregnated surfaces can lead to lubricant migration out of the microstructure[27]. Even after defrosting these surfaces do not recover due to the irreversible lubricant depletion occurring during the frost formation process.[27] Moreover, it has been shown that gravitational force can cause possible drainage of the suffused liquid layer[29] as well. Consequently, loss of the infused lubricant over time occurs, rendering these surfaces effective only for few hours at present. The lack of longevity prevents the direct implementation of the aforementioned surfaces in related industrial applications.

Here, we introduce and demonstrate a novel anti-icing surface concept spontaneously generating and self-preserving a liquid intervening layer through a condensation process. Due to exceptional self-healing properties attributed to the nature of the corresponding infusion mechanism and the stable sustenance of the liquid intervening layer, such surfaces can prolong their operating time under sub-zero conditions significantly, compared to the currently existing surfaces. These type of surfaces are, for example, ideally suitable for integration in ice generation[6,7,39,40] that employs a vacuum freezing process to promote ice nucleation, an integral part of cold thermal energy storage systems (see Supporting Information, section 1).[8,9,41]

## 2. RESULTS AND DISCUSSION

**Working Principle and Realization.** The approach taken here is closely connected to anti-adsorption of water, since the main goal is to form an intervening liquid layer, which hinders water/ice impalement (Figure 1a). First, the interfacial energy of the solid substrate, which in this work is aluminum, is reduced by imparting a random multi-tier texture combined with chemical modification with a hydrophobic coating that consists of perfluorodecyltrichlorosilane (FDTS) and polydimethylsiloxane (PDMS) resulting in high water repellency i.e. advancing contact angle $>150°$ and contact angle hysteresis $<10°$ (Figure 1a top right). Further details regarding the fabrication of the superhydrophobic aluminum-based surface are given in the Experimental

Section. The liquid to be employed for the formation of the intervening layer, apart from its low surface tension, should additionally demonstrate high volatility, immiscibility in water and low freezing point. A liquid ideally suitable for this process is HFE7100, which exhibits low surface tension $\gamma_{HFE}$=13.6 mN/m (25 $^{o}$C), high vapor pressure $P_{HFE,g}$=269.3 mbar (25 $^{o}$C), low freezing point $T_{HFE,fp}$=-135.0 $^{o}$C (1.013 bar) and immiscibility in water. A key feature of HFE7100 is its environmentally friendly characteristic with Global Warming Potential GWP=320, practical nontoxicity and non-inflammability.[42]

To induce the described effects, the textured aluminum substrate, which is cooled and maintained at a temperature within the range $T_{sample}$=[-3.9, -5.2] $^{o}$C, is mounted in a closed chamber which contains water vapor under low pressure at a range of $P_{chamber}$=[180, 280] mbar (see Experimental Section and Supporting Information, section 2). HFE7100 is introduced in the chamber, which results in its evaporation, thus forming a saturated environment that contains a water-HFE7100 mixture. The low temperature of the aluminum surface causes heterogeneous nucleation of HFE7100 on its solid texture, but due to its low freezing point, HFE7100, does not solidify. Attributed to its hierarchically engineered roughness, in conjunction with the very low surface tension of HFE7100, the aluminum surface enables hemi-wicking[43] of the HFE7100, forming a thin and thermodynamically stable intervening layer on the solid substrate due to a constant infusion process based on condensation (Figure 1a). Under these conditions water vapor should also condense on the cooled surface; however, as it has been shown in the literature,[44] it is indicative that HFE7100 hemi-wicking is energetically more favorable compared to water penetration. Moreover, the high water repellency of the aluminum-based surface, in combination

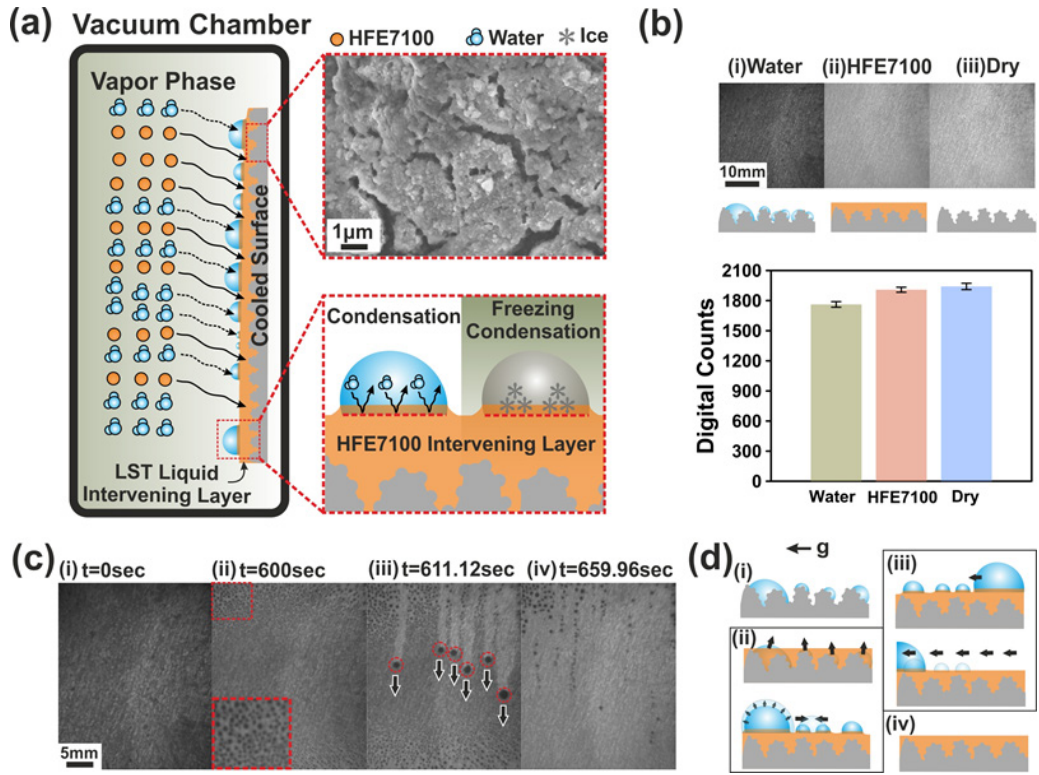

**Figure 1. (a)** Working principle of the surface that involves a cooled aluminum-based superhydrophobic surface (upper right inset) on which condensation-driven impregnation with a low surface tension liquid (HFE7100) occurs, forming a liquid intervening layer. Due to immiscibility of water and HFE7100, water/ice cannot penetrate the asperities of the solid substrate (lower right inset) since the formed intervening layer acts as a barrier to nucleation (condensation and freezing condensation). **(b)** IR images of the target surface being in three different modes: (i) covered with water droplets, (ii) covered with HFE7100 film and (iii) dry. Due to different radiation properties of water, HFE7100 and aluminum-based surface, IR camera receives different magnitudes of average radiation power $\overline{W}_{tot,i} < \overline{W}_{tot,ii} < \overline{W}_{tot,iii}$ (lower panel). **(c)** Mobilization and shedding of impaled water droplets. (i) Initially, at t=0sec the surface is in mode (i). (ii) At t=600sec HFE7100 is introduced in the vacuum chamber rendering

water droplets mobile, resulting in their elevation to the surface of the HFE7100 intervening layer and coalescence with the neighboring drops. (iii) At time instant t=611.12sec droplets have already exceeded the departing size and due to gravitational effect begin to move downwards leaving behind condensate-free tracks. (iv) At t=659.96sec almost the entire surface is freed from water droplets. **(d)** Schematic describing the process of water droplet mobilization and shedding. Direction of gravitational acceleration *g* from left to right.

with the water immiscibility of the formed HFE7100 layer, have a two-fold effect:[37] the condensed water droplets neither diffuse nor submerge in the HFE7100, resulting in the formation of two distinct liquid phases i.e. water floating on the liquid HFE7100 intervening layer. The outcome of this condensation-driven generation of the aluminum-based liquid impregnating surface, is the realization of high slippery characteristics, with low contact angle hysteresis[37] where a condensed droplet resides in a pseudo-Cassie wetting state.[44] Contrary to LIS[27,37,38] and SLIPS[29,36] where the liquid impregnation is occurring in advance in a separate process, for the present case, impregnation is a constantly self-sustained process that replenishes possibly depleted regions, demonstrating a self-healing ability and longevity. The HFE7100 layer acts as a protective barrier against water condensation, and condensate freezing by suppressing icing, considering that in the current study subzero temperature and low pressure occur, as already mentioned previously (Figure 1a bottom right).

**Depinning of Impaled Water Droplets.** Going one step further, in this section we show that HFE7100, due to hemi-wicking effect, can displace water droplets with diameter $d_p \approx 0.4$ mm that have initially impaled the aluminum-based superhydrophobic surface under low pressure and temperature. Similar droplet displacement by HFE7100 has been shown in the literature[44] under air environment and atmospheric pressure. Here mobilization of the water droplets was observed with the use of an infrared (IR) camera which imparts enhanced contrast due to the different radiation properties of condensed water droplets, the HFE7100 intervening layer and the aluminum-based superhydrophobic surface.

To highlight the beneficial effect of using IR for the observations, in Figure 1b, 3 different modes of the aluminum-based superhydrophobic surface are depicted: in mode (i) a small amount of water ($\approx 0.5$ ml) is introduced in the chamber inside which the pressure is maintained at $P_{chamber}=94$ mbar and the temperature at $T_{chamber}=16.7\ ^{o}C$. The temperature of the surface is $T_{sample}=-4.7\ ^{o}C$ and thus water vapor condenses in a form of small droplets on it (Figure 1b(i)). In mode (ii) 150 ml of HFE7100 are inserted in the chamber where pressure and temperature is $P_{chamber}=286$ mbar and $T_{chamber}=17.2\ ^{o}C$ respectively. The sample temperature is maintained at $T_{sample}=-5.0\ ^{o}C$ which results in HFE7100 condensation on the sample surface and the formation of a film (Figure 1b(ii)). In mode (iii), the superhydrophobic surface is kept dry (HFE7100 and water free) at $P_{chamber}=75$ mbar and $T_{chamber}=16.6\ ^{o}C$ with its temperature being kept at $T_{sample}=-5.5\ ^{o}C$ (Figure 1b(iii); see also Supporting Table S1 for a summary of the pressure and temperature values obtained and Table S2 for the estimation of the measurement uncertainty). It should be noted that the total radiation power received by the IR camera sensor $W_{tot}$, measured in digital counts, scales with the emitted power of the superhydrophobic surface $W_{SH}$ as $W_{tot} \propto (1-\alpha_j)\varepsilon_{SH} W_{SH}$ [45,46] where $\varepsilon_{SH}$ is the emissivity of the superhydrophobic surface, $\alpha_j$ the absorbtivity of water ($j \equiv i$, mode (i)) or HFE7100 ($j \equiv ii$, mode (ii)). For mode (iii) $j \equiv iii$ that corresponds to dry superhydrophobic surface $\alpha_{iii}=0$ which yields $W_{tot} \propto \varepsilon_{SH} W_{SH}$ (see Supporting Information, section 3). Absorptivity of water $\alpha_i \approx 0.98$ [47] is greater than that of HFE7100 $\alpha_{ii} \approx 0.19$ [48] (see Supporting Information, section 3). Thus considering that $1-\alpha_i < 1-\alpha_{ii} < 1-\alpha_{iii}$ the respective total radiation power follows the same trend $W_{tot,i} < W_{tot,ii} < W_{tot,iii}$ as shown in the column graph of Figure 1b in which the average total radiation power $\overline{W}_{tot,j}$ is calculated, received by all the individual detectors of the IR camera's focal plane array. This deviation in the received total radiation power for the aforementioned 3 cases is reflected at the intensity of the acquired IR images as shown in Figure 1b and high contrast between water droplets and HFE7100/superhydrophobic surface can be achieved.

In order to show the impact of HFE7100 hemi-wicking on droplet mobilization, 600 seconds after the onset of mode (i) (Figure 1b and Figure 1c(i)), a volume of 150ml HFE7100 is introduced in the vacuum chamber and HFE7100 condenses. Due to its low surface tension ($\gamma_{HFE}=13.6$ mN/m at $25^{\circ}C$) compared to water ($\gamma_W=72.0$ mN/m at $25^{\circ}C$) in combination with the high water repellency of the cooled surface, the impaled water droplets become depinned and are elevated to the surface of the formed HFE7100 intervening layer (Figure 1c(ii) and 1d(ii)). Attributed to their mobility restoration, the droplets coalesce with adjacent ones resulting in their growth (Figure 1c(ii) and 1d(ii)). Within a short period of time (<11.12sec), after reaching a critical diameter of approximately $0.4$ mm where the gravitational forces exceed the capillary forces, the droplets depart and move along vertical paths (Figure 1c(iii) and 1d(iii)). During the downward motion, the droplets wipe the surface clean from water leaving water condensate-free traces, suggesting that the latter are covered only by HFE7100. The formation of condensate-free paths is indicated by their bright color, which is associated with the presence of only HFE7100. The outcome of HFE7100 hemi-wicking is that 59.96sec after its insertion in the vacuum chamber the cooled superhydrophobic surface has already become water-free (Figure 1c(iv) and 1d(iv)). By identifying droplet paths on the surface area (see Supporting Information, section 4) and based on the corresponding measured $\overline{W}_{tot}$ it is possible to determine the occurrence of droplet mobilization and shedding. Initially, due to the coverage of a path by water droplets, $\overline{W}_{tot}$ has values (in digital counts) ranging between [1790, 1820]. With the mobilization initiation of a droplet in the path, $\overline{W}_{tot}$ increases abruptly with time to [1850, 1860], since the moving droplet is fully swiping the path from water condensate. After the occurrence of a mobility event, $\overline{W}_{tot}$ drops gradually vs time since the water-free path is progressively covered with new condensate (see Supporting Information, section 4 and Supporting video S1).

**Condensate Droplet Mobility.** Next we investigate the droplet mobility of condensate that is formed on the surface under steady operation conditions. Based on image sequences acquired with the IR camera it is feasible to investigate the droplet mobility during the condensation process of water vapor on the self-impregnating surface (see Supporting video S2). For that 60ml of water and 150ml of HFE7100 are introduced in the vacuum chamber implementing the procedure described in the Supporting Information, section 2 with sample temperature, chamber temperature and pressure being kept at $T_{sample}$=-3.5 °C, $P_{chamber}$=238 mbar and $T_{chamber}$=11.6 °C, respectively (see Table S2 for the estimation of the measurement uncertainty). As described in the Supporting Information section 4, 6 main paths are visible on the surfaces (Figure 2a) the shape of which is the trace of a droplet that begins to move vertically down the surface at a specific time instant during the mobility study duration (300sec). Within this time a number of additional droplets enter these main paths and others may form inside the paths and begin to move downwards. Consequently, a time series of $\overline{W}_{tot}$ can be plotted for each main pathway as shown in Figure 2b. It should be noted that each time series and its corresponding path have the same plot and contour color respectively. Time series in black color corresponds to $\overline{W}_{tot}$ of the

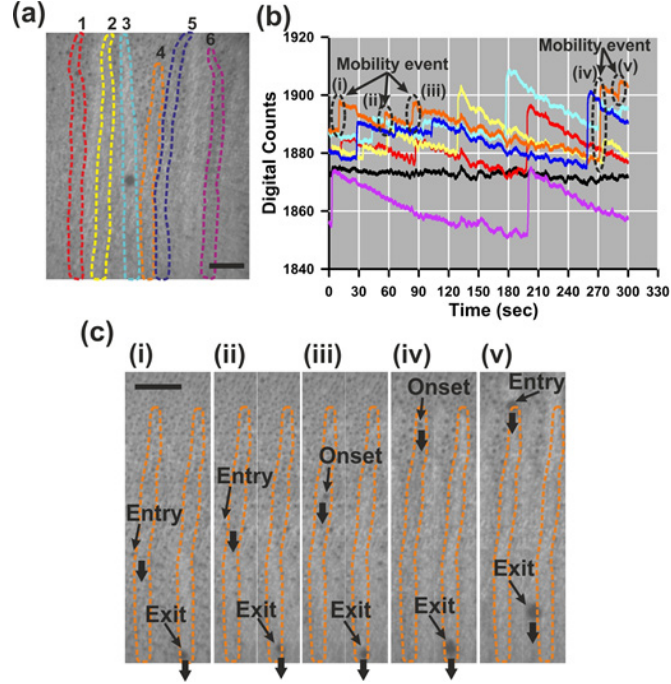

**Figure 2.** **(a)** Drawing of 6 main paths for the study of the droplet mobility on the self-impregnating surface. **(b)** Time series of $\bar{W}_{tot}$ (in digital counts) for each main path. Each pair of local minima and maxima correspond to a mobility event ((i)-(v)). **(c)** Mobility events (i)-(v) occurring at main pathway No 4 (orange contour). Droplets enter or form inside the path, fully or partially removing the condensate. Scale bars correspond to 5mm.

entire window. Each series exhibits a number of local maxima ($\bar{W}_{tot,max}$) and minima ($\bar{W}_{tot,min}$), which are associated with the onset and completion of a mobility event. The latter involves the entrance or mobilization onset of a droplet in a path at a specific time instant (Figure 2c). The magnitude of the abrupt change of $\bar{W}_{tot}$ ($\Delta\bar{W}_{tot}=\bar{W}_{tot,max}-\bar{W}_{tot,min}$) depends on whether the condensate covering the main path is fully or partially removed by a moving droplet. Considering main pathway No 4 (orange contour), 5 mobility events are identified (Figure 2c):

in mobility events (i), (ii) and (v) a droplet enters a path from outside at a certain point and partially removes water condensate resulting in an abrupt change of small magnitude ( $\Delta \bar{W}_{tot} \approx 3\text{-}10 \text{ digital counts}$ ). In mobility event (iii) a droplet is formed inside the path and mobilized at a certain distance from the starting point of the path. Mobility event (iv) is similar to (iii) with the difference that the droplet initiation takes place at a very beginning of the corresponding path. It should be noted that the shape of all paths was defined based on mobility event (iv). Based on the above, an abrupt signal increase of $\bar{W}_{tot}$, much greater compared to the rest of the events ( $\Delta \bar{W}_{tot} \approx 25 \text{ digital counts}$ ), is observed when event (iv) occurs. The time-series of the overall $\bar{W}_{tot}$ (black color), shows that it is approximately constant with time which indicates there is an equilibrium between the number of droplets that shed and the ones that form.

Droplet mobility is one of the integral factors of ice nucleation delay due to the fact that droplets do not stick on the present surfaces, which limits their contact time with the cooling surface and restricts the probability of freezing.

**Freezing Condensation Delay.** To highlight the superior performance of the present surface design i.e. the self-impregnating superhydrophobic aluminum-based surface, its anti-icing behavior was compared (Figure 3a, lower frame sequence) to an untreated smooth aluminum sample (Figure 3a, upper frame sequence) using a liquid mixture of 60ml/150ml of water/HFE7100 as described in the previous section. Moreover, the performance of the aluminum-based superhydrophobic surface and the untreated smooth aluminum were tested under only water vapor exposure as well. This was materialized by applying the superhydrophobic fabrication technique described in the Experimental Section to the half of the area of a circular smooth aluminum sample leaving the other half untreated; thus two hemicircular regions were formed (Figure 3b, left and right corresponding to superhydrophobic and untreated surface respectively). To ensure same reference time for all cases, start of study (Figure 3b and 3c, t=0min) corresponds to the time instant that the sample temperature becomes $T_{sample}$=0 °C. At this time instant the gaseous phase temperature inside the chamber is $T_{chamber}$=17.4 °C and the corresponding pressure for the case where the chamber contains HFE7100/water vapor is $P_{chamber}$=288 mbar. For the case of only water vapor, $P_{chamber}$=95 mbar. In addition, supersaturation $S^{11,49}$ is calculated S=3.25, defined as $S=P_{w,\,vapor}/P_{w,\,sat}(T_{sample})$ where $P_{w,\,vapor}$ is the pressure or the partial pressure (for the case of water/HFE7100 mixture) of water in the chamber and $P_{w,\,sat}$ is the saturation pressure of water at $T_{sample}$. It should be noted that the temperature setpoint of the chiller is $T_{chiller}$=-13.0 °C.

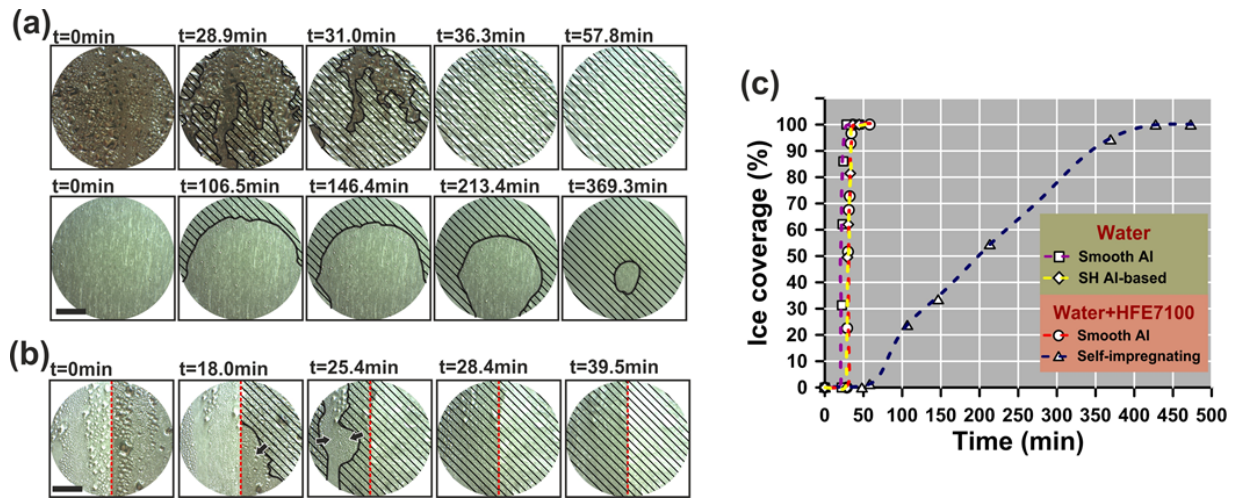

**Figure 3. (a)** Time lapse images of ice nucleation showing onset and propagation of icing on a smooth aluminum surface (upper frame sequence) and the present surface concept namely the self-impregnating aluminum-based superhydrophobic surface (lower frame sequence). **(b)** Image sequence depicting onset and development of icing on aluminum superhydrophobic (left hemicircular disc zone) and smooth (right hemicircular disc zone) surface under only water vapor exposure. **(c)** Ice coverage vs time for smooth aluminium (Smooth Al) and superhydrophobic aluminum-based (SH Al-based) surface under water vapor and water/HFE7100 mixture exposure. For water/HFE7100 mixture, superhydrophobic aluminum-based surface becomes self-impregnating. t=0min is the time instant that $T_{sample}$=0 °C for all the examined cases. Scale bar corresponds to 10mm. Regions covered with ice are indicated with hatching.

During exposition to water/HFE7100 mixture and due to the hydrophilic property of the smooth untreated aluminum sample, cumbersome and random-shaped water condensate droplets cover its entire surface. After a lapse of 28.0min with $T_{chamber}$=16.8 °C, $T_{sample}$=-3.3 °C, $P_{chamber}$=282 mbar and S=3.99, onset of freezing is observed. Nucleation is triggered primarily from the edge of the sample where presumably realistic surface impurities and geometric characteristics (due to fabrication and handling process) lower the free energy barrier for heterogeneous nucleation $\Delta G_{het}^{smooth}$ compared to the internal part of the sample.[10,50–54] The beginning of freezing was identified due to change of the opacity of the frozen water droplets. This edge mode nucleation[54] initiates a freezing front which propagates inwards along the interconnected water droplets that are pinned on the test surface[55] (Figure 3a, upper frame sequence, t=28.9, 31.0min). The propagation of this front is relatively fast and within 7.9min after the freezing initiation, with $T_{chamber}=16.4\,^{\circ}C$, $T_{sample}=-4.4\,^{\circ}C$, $P_{chamber}=278\,mbar$ and $S=4.23$, the surface is fully covered with frozen condensate (Figure 3a, upper frame sequence, t=36.3min and Figure 3c) and an ice layer with growing thickness is developing (Figure 3a, upper frame sequence, t=57.8min).

Contrary to the aluminum smooth surface, the self-impregnating surface at t=0min is covered by a small amount of droplets with diameter $\leq 0.4\,mm$. As discussed previously even at this small scale the droplets are mobile and finally shed from the surface. This unique characteristic retards the propagation of a formed freezing front (Figure 3a, lower frame sequence, t=0min). It should be noted that a self-impregnating surface due to its HFE7100 intervening layer demonstrates a molecular scale smooth interface[29] which brings the energy barrier for heterogeneous nucleation $\Delta G_{het}^{self-impregnating}$ much closer to that of homogeneous nucleation $\Delta G_{hom}$. Consequently, this indicates that $\Delta G_{het}^{self-impregnating} > \Delta G_{het}^{smooth}$ [54] implying that ice nucleation will occur easier on a smooth solid surface than on a liquid intervening layer of self- impregnating surface (see Supporting Information, section 5 for derivation of heterogeneous nucleation on smooth aluminum and self-impregnating surface). Indeed, at time $t \approx 58.0min$ with $T_{chamber}=16.4\,^{\circ}C$, $T_{sample}=-4.7\,^{\circ}C$, $P_{chamber}=278\,mbar$ and $S=4.33$, ice formation is observed at the edges as in the case of the smooth aluminum with a delay of 30.0min. Likewise, the presence of HFE7100-free regions at the edges of the test sample lowers the $\Delta G_{het}^{self-impregnating}$ increasing the potential for ice nucleation. However, the freezing region that is generated from this edge mode nucleation propagates very slowly inwards where the interface of the self-impregnating superhydrophobic substrate consists of the slippery HFE7100 intermediate layer (Figure 3a, lower frame sequence, t=106.5, 146.4, 213.4, 369.3min). It seems that the front of this region is assisted by partially migration of HFE7100[27] to the surface of the icing front causing local HFE7100 depletion that exposes the solid substrate to condensing water. This is attributed to the nanoscale roughness of the ice front generated by the formed ice crystals that induces development of capillary forces causing attraction of HFE7100.[27] However, since there is a continuous process of HFE7100 condensation on the substrate as well, it is speculated that there is a competition process between HFE7100 migration from the substrate to the formed ice and HFE7100 condensation which finally results in exposing only the roughness peaks to water vapor. Therefore, we believe that the occurrence of condensation restricts the impact of HFE7100 migration and retards significantly irreversible lubricant depletion that takes place to liquid presuffused surfaces.[27,29] After a remarkable lapse of 428.0min with $T_{chamber}$=16.8 $^{o}$C, $T_{sample}$=-5.1 $^{o}$C, $P_{chamber}$=282 mbar and S=4.58, self-impregnating surfaces have been completely covered with ice (Figure 3c) which underpins its extreme anti-icing behavior. It should be noted that during the partial icing, the remaining ice-free region retains its highly slippery characteristics.

The superhydrophobic sample exposed only to water vapor (Figure 3b) demonstrated a behavior similar to the smooth sample which was exposed to water/HFE7100 vapor mixture. Ice nucleation was initiated in both hemicircular disc zones, namely the smooth and the superhydrophobic, by the test sample edge followed by a freezing propagation front (Figure 3b t=18.0 and 25.4min). As expected, the smooth zone exhibited onset of icing earlier ($t \approx 17.0$min, $T_{chamber}$=17.2 $^{o}$C, $T_{sample}$=-2.6 $^{o}$C, $P_{chamber}$=89 mbar and S=3.88) compared to the superhydrophobic zone ($t \approx 24.0$min, $T_{chamber}$=17.0 $^{o}$C, $T_{sample}$=-2.9 $^{o}$C, $P_{chamber}$=97 mbar and S=3.93) with the 100% freezing coverage been accomplished at $t \approx 23.0$min ($T_{chamber}$=17.0 $^{o}$C, $T_{sample}$=-2.8 °C, $P_{chamber}$=96 mbar and S=3.91) and $t \approx 31.0 min$ ($T_{chamber}$=16.8 °C, $T_{sample}$=-3.3 °C, $P_{chamber}$=92 mbar and S=3.98) respectively (Figure 3b t=28.4 and 39.5min, after the complete ice coverage of smooth and superhydrophobic surface respectively; see also Supporting Table S1 for a summary of the pressure and temperature values obtained and Table S2 for the estimation of the measurement uncertainty). Despite the fact that the superhydrophobic surface is supposed to promote high rolling characteristics, the departure droplet size is considerably higher $d_p \approx 2.7$ mm compared to the case of the self-impregnating surface. It is noteworthy that even though superhydrophobic surfaces are connected to anti-icing,[14,56] in the present study the aluminum-based superhydrophobic surface does not have exceptional icephobic behavior since it delays ice nucleation by only 7min compared to the smooth aluminum sample. Similar trend is verified with a different chiller temperature setpoint of $T_{chiller}$=-30.0 °C where the aluminum-based superhydrophobic surface delays ice nucleation by only 3min (see also Supporting Information, Table S3).

Figure 3c highlights the anti-icing potential of all the cases studied. It is obvious that the self-impregnating surface has superior performance compared to the smooth and superhydrophobic surface with a full coverage occurring within a period of one order of magnitude, 10-15 times, longer than that of the rest of the cases. Moreover, the onset of icing in the self-impregnating surface occurs after a period which is 2-3.5 times greater than that of aluminum smooth (with water/HFE7100 mixture or only water) and superhydrophobic samples.

**Skating Ice During Defrost Process.** Next we study the defrosting behavior of self-impregnating surfaces, important for removal of potentially accumulated ice from a heat exchange interface (e.g. in ice production systems). The sample is exposed to subcooling until it reaches $T_{sample}$=-9.8 °C, $T_{chamber}$=13.7 °C and $P_{chamber}$=255 mbar. At this point an ice layer has been built covering the entire surface area and cooling is interrupted raising $T_{sample}$ with a rate of approximately 0.3 °C/min. The onset of melting is identified with the use of an IR camera (t=0sec). It should be noted that ice, similarly to water, has an absorptivity of $\alpha_{ice} \approx 0.98$ [47] and thus negligible transmissivity $\tau_{ice}$ resulting in high contrast in the acquired images between frozen (black color) and ice–free regions (grey shade). At t=53.32sec (Figure 4a) melting is expanding spatially and at t=72.24sec segments of ice progressively detach from the parent ice layer and finally start skating on the self-impregnating surface (Figure 4a, t=73.96, 75.60sec) leaving approximately 90% of the surface area clean from ice (see Supporting Video S3). This indicates that ice adhesion on the self-impregnating surface is much lower compared to the rest of the cases where no skating of ice was observed. At $t \approx 167.00$sec ice has been totally removed from the self- impregnating surface ($T_{sample}$=-0.1 °C, $T_{chamber}$=13.9 °C and $P_{chamber}$=256 mbar; see also Supporting Table S1 for a summary of the pressure and temperature values obtained and Table S2 for the measurement uncertainty estimation). After completion of ice melting, the aluminum smooth and superhydrophobic surface exposed only to water vapor (Figure 4b, left and right hemicircular disc zones respectively), are covered with pinned water condensates and the smooth aluminum sample exposed to water/HFE7100 is covered with large regions of collected water (Figure 4b, right hemicircular disc zone) whereas the self-impregnating surface is free of water droplets (Figure 4b, left hemicircular disc zone).

A possible explanation of the skating behavior of ice is presented in Figure 4c. As ice starts to form at the edge of the sample due to the presence of surface defects (Figure 4c(i)), it expands inwards (Figure 4c(ii)) with simultaneous partial HFE7100 migration that leaves the roughness peaks exposed to icing (Figure 4c(iii)). After interruption of cooling a thin water layer is formed at the interface of ice-HFE7100-aluminum sample (Figure 4c(iv)). However, as already discussed, since the aluminum substrate has superhydrophobic properties and HFE7100 has much lower surface tension compared to water, the liquid layer that wets the peaks of the surface roughness is removed by the remaining HFE7100 (Figure 4c(iv)). This will result in the formation of a slippery thin layer of HFE7100 underneath the ice layer (Figure 4c(v)) leading finally to shedding of ice that resembles skating (Figure 4c(vi)). Sliding of ice on presuffused surfaces has been reported in the literature;[29] however, to our knowledge, the unique advantages of observation in the IR spectrum, to distinctly show and verify the occurrence of ice skating were not exploited before. In addition, concerning the durability of pre-impregnated surfaces,[27,29] it has been shown that after defrosting cycles degradation of their performance occurs due to the fact that part of the suffusing low surface tension liquid is being depleted.[54]

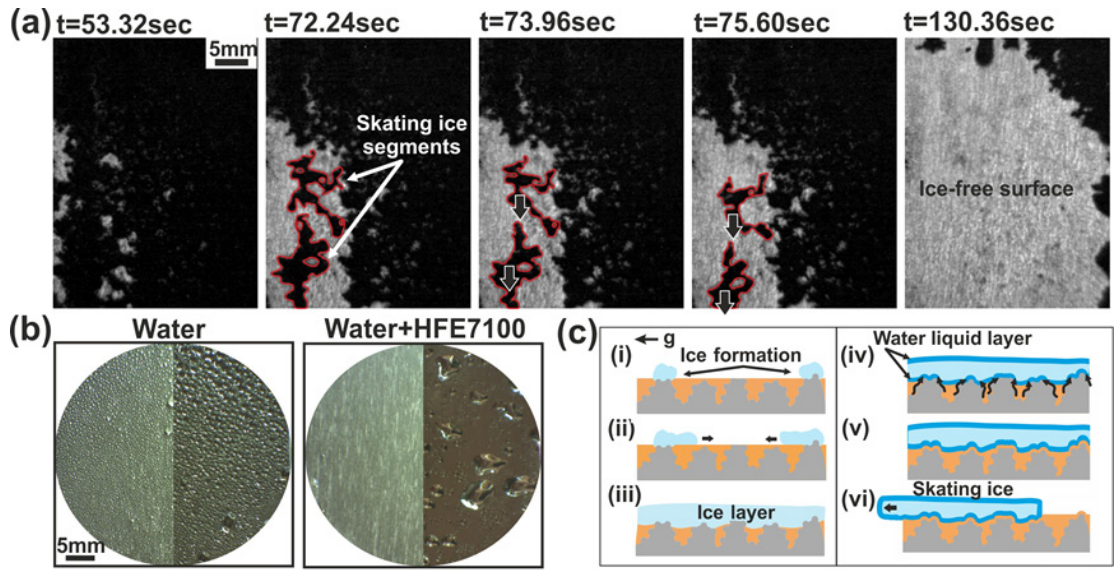

**Figure 4. (a)** Time lapse IR images of defrosting process on self-impregnating slippery surfaces. t=0sec corresponds to the time when melting was apparent through IR observation. **(b)** Images in the visible spectrum (CMOS) of the tested cases after the end of the defrosting process. Left panel: SH (left hemicircular disc) and smooth aluminum (right hemicircular disc) surface under exposure only to water vapor. Right panel: Superhydrophobic self-impregnating (left hemicircular disc) and smooth aluminum (right hemicircular disc) surface under exposure to water/HFE7100 vapor mixture. **(c)** Schematic showing a proposed mechanism of skating of ice segments on self-impregnating surface. Direction of gravitational acceleration $g$ from right to left.

**Longevity.** Robustness is an integral feature of an engineered surface especially if it is intended for implementation in real applications. Therefore to ensure sustainability of functionality for use in ice production systems a surface has to withstand prolonged subcooling and vacuum conditions. To this end we performed a series of experiments that involved long-term operation of the self-impregnating surface at $T_{sample}$=-3.9 °C, $T_{chamber}$=0.7°C, $P_{chamber}$=179 mbar and $S$=1.40 for a total duration of 490hr (see Table S2 for the measurement uncertainty estimation of pressure and temperature). At times t=0, 200, 300, 400, 490 hr the operation was interrupted and measurements of advancing ($\theta_A$) and receding ($\theta_R$) contact angles of the dry aluminum-based superhydrophobic surface were acquired (Figure 5a and 5b, sample I). Changes in wettability which is characterized through measuring the contact angles is closely related to the chemical and/or mechanical degradation of the test surface and therefore is used extensively in research works for robustness characterization[57,58]. Additionally, two aluminum-based superhydrophobic surfaces were dipped in HFE7100 (Figure 5b, sample II) and water bath (Figure 5b, sample III) each at room temperature, followed by advancing and receding contact angle measurements after 807hr and 30hr respectively. The aluminum-based superhydrophobic substrate, which is also a component of the self-impregnating surface, initially exhibits $\theta_A=165.0°$, $\theta_R=157.2°$ and contact angle hysteresis $\theta_H=\theta_A-\theta_R=7.8°$. After a long lapse of 490hr superhydrophobic sample shows remarkably only a slight degradation of its surface with a decrease of 3.0 and 6.1 degrees in its advancing and receding contact angle respectively, whereas its contact angle hysteresis increased by 3.1 degrees (Figure 5a). This indicates that the self-impregnating surface practically retains its full functionality for at least 490hr since the experiments did not continue further. This exceptional performance is attributed to the beneficial presence of the HFE7100 intervening layer that does not interact chemically with the superhydrophobic component of the self-impregnating surface, preventing hydrolyzation and thus degradation of the hydrophobic coating by water. This is verified by the performance of samples II and III. Indeed after 807hr of exposure in a HFE7100 bath, sample II, which initially exhibits $\theta_A=166.9°$, $\theta_R=162.9°$ and $\theta_H=4.0°$, shows a drop of 2.1 and 1.4 degrees in its advancing and receding contact angle respectively giving an increase of 0.7 degrees in its contact angle hysteresis. Therefore it is apparent that HFE7100 is effectively inert with respect to the superhydrophobic texture. Conversely, sample III (initial contact angles: $\theta_A=166.4°$, $\theta_R=159.8°$ and $\theta_H=4.5°$) demonstrates a decrease in advancing and receding contact angles by 4.5 degrees and 27.7 degrees, respectively, resulting in a considerable increase in the contact angle hysteresis of 23.1 degrees and indicating the water degenerative action over the surface.

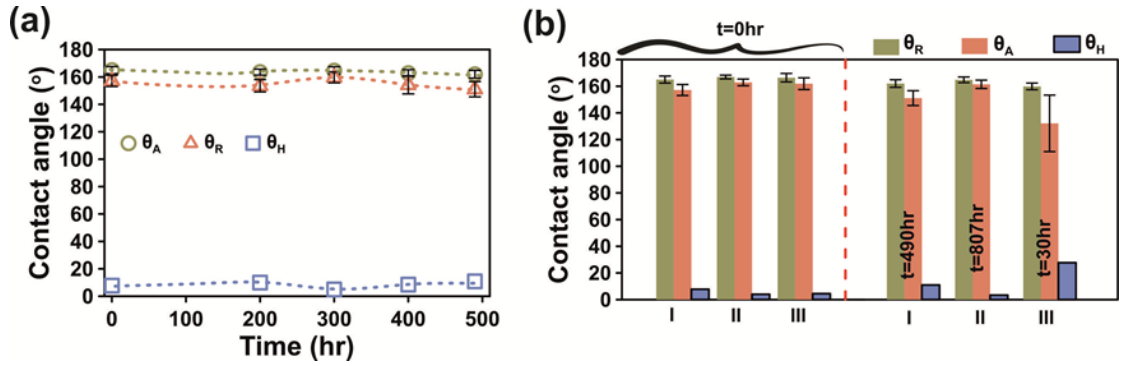

**Figure 5. (a)** Longevity tests on self-impregnating surface; advancing $\theta_A$, receding $\theta_R$ and hysteresis contact angle vs time at $T_{sample}=-3.9°C$, $T_{chamber}=0.7°C$, $P_{chamber}=179$ mbar and $S=1.40$ up to 490hr. **(b)** Advancing, receding $\theta_A$, $\theta_R$ and contact angle hysteresis $\theta_H=\theta_A-\theta_R$ corresponding to three different stability tests before and after the total exposure; sample I is the one used for the longevity tests described in (a), sample II was dipped in HFE7100 for 807hr and sample III was exposed to water for 30hr. Error bars represent the standard deviation of the contact angle measurements.

## 3. CONCLUSION

An innovative anti-icing strategy was introduced that involves a new and prospective concept of self-impregnating slippery surfaces. With a rational choice of materials and applying wettability engineering fundamentals, these surfaces have unique properties suitable for real applications such as ice generation. It was shown that, due to their slippery behavior, these surfaces suppress ice nucleation. To observe the performance, we use an IR camera to follow the droplet depinning by HFE7100 hemi-wicking. Exploiting the different radiation properties of water, HFE7100 and aluminum-based surfaces, the complex mechanism of droplet shedding was studied which is not visible by using a CMOS camera. To this end, a cycle of microscale droplet growth, coalescence and shedding was revealed, that wipes the water condensates leading to formation of vertical water-free paths which comprise only HFE7100. In addition, droplets with a diameter of only $d_p \approx 0.4$ mm were rendered mobile in contrast to the smooth (pinned water condensate) and aluminum-based superhydrophobic surface (departing droplet size there was $d_p \approx 2.7$ mm). Due to their slippery characteristics, molecular-scale smooth interface and self-generating capability originating from the continuous condensation process, the self-impregnating surfaces demonstrated an exceptional anti-icing behavior under subcooling and vacuum conditions. They remarkably delayed the onset of icing by 2-3 times and the complete ice coverage by 10-15 times compared to smooth and superhydrophobic surfaces exposed to water vapor only. During defrosting ice segments were skating before finally shedding, leaving the surface ice-free. The self-impregnating slippery surfaces exhibited excellent stability under subcooling and vacuum conditions. After a lapse time of 490hr the self-impregnating surface retained its functionality without marked chemical and mechanical degradation. Taken together, the aforementioned properties render this anti-icing approach a highly prospective strategy for direct implementation in industry and especially in ice production systems.

## 4. EXPERIMENTAL SECTION

**Fabrication of Aluminum-based Superhydrophobic Surface.** The fabrication process of the superhydrophobic texture is described elsewhere.[44,59] Briefly, a circular mirror-like polished aluminum substrate (AW 1085) with diameter 60 mm and thickness of 1mm was dipped in a sodium hydroxide (NaOH) aqueous solution (1% w/w) in order to remove the native oxide layer that has formed on its surface. Subsequently, the sample was immersed in a ferric chloride ($FeCl_3$) aqueous solution (1M) followed by dipping in a hydrogen peroxide ($H_2O_2$) aqueous solution (30% w/w). After this step the surface exhibits superhydrophilic properties and micro/nano textured morphology (Figure 1a top right). At a second step the substrate is functionalized with immersion in a perfluorodecyltrichlorosilane (FDTS) – hexane solution (0.05% v/v). Consequently, the sample became superhydrophobic with advancing contact angles >150º and contact angle hysteresis <10º. To enhance durability, the hydrophobic aluminum samples were dip-coated in a polydimethylsiloxane (PDMS, 50 mg, 10% curing agent)–FDTS (100 μL)–THF (Tetrahydrofuran, 10 mL) solution followed by curing in the oven. The measured apparent advancing $\theta_A$ and receding $\theta_R$ contact angles of a $10\,\mu L$ water droplet deposited on the final surface is 166.1° and 160.6°, respectively.

**Wettability Characterization.** Wettability of the surfaces was evaluated by conducting measurements of advancing and receding contact angles with the employment of the dynamic sessile drop technique using a home-built goniometer. Using a needle, a microliter-scale sessile water drop is deposited on the surface on which is allowed to expand and contract. This expansion and contraction process of the droplet was controlled with a syringe pump (New Era, NE1000-E). Images of the dynamic change of the droplet volume were captured with backlit image acquisition setup, which consisted of a camera (Thorlabs DCC1545M, CMOS) fitted with a zoom lens (Thorlabs Zoom 7000 TV Lens MVL7000). Images were processed with ImageJ to finally calculate the contact angles. 3-10 contact angle measurements (advancing and receding) were acquired by placing a water drop at different spots of the surface each time. Inspection of the topography of the surfaces was materialized with a Zeiss ULTRA 55 scanning electron microscope (SEM). Due to the electrically non-conductive property of the aluminum-based samples, prior to characterization, these samples were sputter coated with a layer of Pt to prevent charge accumulation which resulted in image distortion.

**Experimental Setup for Anti-icing, Defrosting and Longevity Observation.** A detailed description of the apparatus can be found in the Supporting Information, section 2. Briefly, the circular aluminum samples were mounted on the planar side of a cylindrical aluminum block inside a vacuum chamber. The other planar side of the block, is in contact with the cooled side of the chamber ensuring cooling of the test surface. The block lateral wall was thermally insulated to ensure minimization of the heat losses through it and one-dimensional heat conduction. Along the axis of the block, thermocouples are mounted in order to calculate the heat flux (see Supporting Information, Figure S2). A thermocouple and a pressure sensor are also mounted inside the chamber to accommodate measurements of the vapor phase temperature and pressure, respectively. Initially, while being cooled, the chamber is evacuated to approximately 85mbar to minimize the presence of non-condensable gases from the chamber, followed by introducing water and subsequently HFE7100. The outcome of this procedure is the formation of a liquid and gaseous phase of the binary mixture i.e. water-HF7100. A sapphire glass is mounted at one of the planar sides of the chamber, which enables observation of the target sample optically (CMOS camera) and in the infrared (IR) spectrum. The use of IR camera is advantageous compared to the CMOS camera since it provides considerably enhanced contrast between water/ice covered regions and regions where the superhydrophobic surface is covered only with HFE7100. As it has been already discussed in Results and Discussion section this is attributed to the different emissivities of water, ice, HFE7100 and aluminum.

## ASSOCIATED CONTENT

**Supporting Information.** The Supporting Information is available free of charge on the ACS Publications website at DOI

Contents: Electro-thermal energy storage (ETES) system, experimental apparatus, observation at the infrared (IR) spectrum, IR observation of impaled droplet depinning on self-impregnating surface, heterogeneous nucleation on self-impregnating surface vs smooth aluminum surface, supporting figures (S1-S4), supporting tables (S1-S3), supporting videos (S1-S3).

**AUTHOR INFORMATION**

**Corresponding Author**

*E-mail: dpoulikakos@ethz.ch. Phone: +41 44 63 22738. Fax: +41 44 63 21176 (D.P.)

**Author Contributions**

D.P. and C.S. conceived the idea to perform this research. D.P., J.H. and C.S. guided the research scientifically and technically throughout. C.S., J.H. and D.W. performed all experiments and analyzed data. C.S., D.P. and J.H. wrote the paper. All authors discussed the content of this work and reviewed the manuscript. The manuscript was written through contributions of all authors. All authors have given approval to the final version of the manuscript.

**Notes**

Authors declare no competing financial interests.

**ACKNOWLEDGMENTS**

C.S. and D.P. acknowledge partial support from an industrial grant (ABB). D.P. acknowledges partial support from the European Research Council (ERC) Advanced Grant 669908 (INTICE).

The authors would like to thank Dr. Manish Tiwari and Dr. Hadi Eghlidi for their insightful comments within the context of this work and Bruno Krammer for his technical assistance.

SYNOPSIS (TOC)

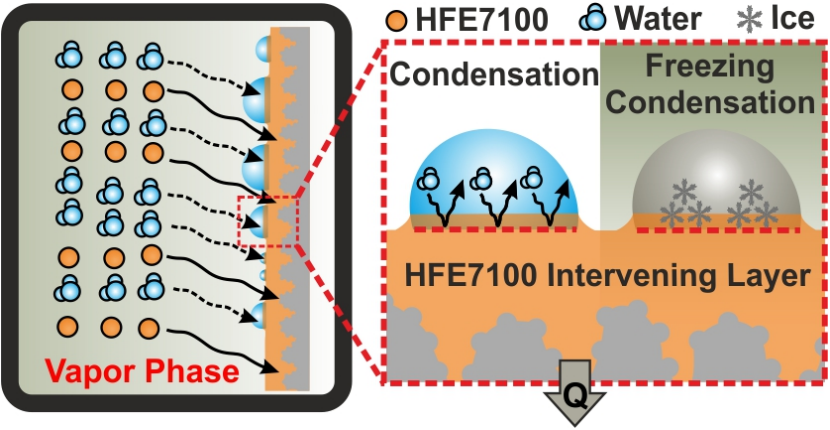